# Ordinateur porté support de réalité augmentée pour des activités de maintenance et de dépannage

Olivier CHAMPALLE, Bertrand DAVID, René CHALON, Guillaume MASSEREY

Laboratoire ICTT, Ecole Centrale de Lyon

36, avenue Guy de Collongue, 69134 ECULLY cedex +33 4 72 18 65 81

{Olivier.Champalle, Bertrand.David, Rene.Chalon, Guillaume.Masserey}@ec-lyon.fr

## RESUME

Nous présentons dans cet article une étude de cas d'utilisation d'ordinateur porté dans le cadre d'activités de maintenance et de dépannage. Outre la problématique de configuration de cet ordinateur et des périphériques à lui adjoindre, nous montrons l'intégration des aspects contexte, stockage in-situ, traçabilité et prescription d'opérations dans ces activités. Cette étude de cas s'inscrit dans le cadre d'un projet plus vaste appelé HMTD (Help Me To Do) qui étudie l'utilisation des principes MOCOCO (MObilité, COoperation, COntextualisation) et IMERA (Interaction Mobile dans l'Environnement Réel Augmenté) pour une meilleure utilisation, maintenance et dépannage d'équipements dans les contextes domestique, public et professionnel.

## Mots Clés

Interfaces Mobiles, Environnements de Réalité Mixte, Travail collaboratif, Intelligence ambiante.

## ABSTRACT

In this paper we present a case study of use of wearable computer within the framework of activities of maintenance and repairing. Besides the study of configuration of this wearable computer and its peripherals, we show the integration of context, in-situ storage, traceability and regulation in these activities. This case study is in the scope of a huge project called HMTD (Help Me To Do) which aim is to apply MOCOCO (Mobility, COoperation, COntextualisation) and IMERA (Mobile Interaction in the Augmented Real Environment) principles for better use, maintenance and repairing of equipments in the domestic, public and professional situations.

## Categories and Subject Descriptors

D2.2 [Sofware Engineering]: Design Tools and Techniques; H5.2 [Information Interfaces and Presentation]: User Interfaces.

## 1. INTRODUCTION

Comme annoncé par Weiser [12], l'informatique ubiquitaire est en plein essor et semble se concrétiser avec la propagation massive des dispositifs mobiles et connectés (TabletPC, PDA, Smartphones,…). Ces ordinateurs portés soulèvent bien des enjeux [11] que Mann [8] essaya de caractériser en « 6 flux de signaux de bases » tels que la « non monopolisation de l'attention de l'utilisateur » ou encore la « prise en compte de l'environnement ». Notre cadre d'étude étant la maintenance [6], nous cherchons notamment à intégrer ces ordinateurs portés dans des applications d'assistance où un technicien est envoyé sur site d'intervention avec un système mobile supportant la collaboration [1, 8], présentant des données contextualisées [1] et pouvant mettre en œuvre des dispositifs de réalité augmentée (RA) [6].

Notre objectif est donc d'étudier l'informatique mobile, portée [11] et ubiquitaire dans une orientation de réalité augmentée [13], et proposer des interfaces hommes machines innovantes appropriées pour les utilisateurs mobiles travaillant d'une manière collaborative dans des environnements précis nécessitant l'accès aux données contextualisées dans une logique de réalité augmentée, grâce, notamment, à la maturité de la technologie RF-ID. Nos principaux concepts sont :

- IM (Interfaces utilisateur Mobiles), il décrit les interfaces utilisateur pour PDA, Smartphone et d'autres dispositifs adaptés aux situations mobiles,
- ERA (Environnement Réel Augmenté) dans le sens de la réalité mixte et l'informatique ubiquitaire,
- MOCOCO (MObilité, COoperation, COntextualisation) décrit les tâches réalisées en collaboration par plusieurs acteurs mobiles, qui ont accès à des données précises et contextualisées
- Proactivité, l'anticipation de l'interface par rapport aux actions utilisateur dans une logique d'intelligence ambiante.

La plateforme IMERA sert de support d'intégration de ces différents principes et permet l'adaptation aux différentes situations applicatives.

Dans la suite de cet article nous décrivons sommairement la plateforme IMERA et le projet HMTD qui prend en compte quatre aspects importants qui sont la notion de contexte, celle de stockage d'informations in-situ, de traçabilité et de prescription d'opérations. Puis nous présentons un scénario concret de dépannage industriel, nous en décrivons les tâches correspondantes et la configuration de l'ordinateur porté à laquelle nous avons abouti. Enfin, nous utilisons le formalisme IRVO [2] pour décrire et comparer les interactions de RA, avant de conclure.

## 2. Plateforme IMERA

La plateforme IMERA, déployée sur le campus de l'Ecole Centrale de Lyon, s'appuie sur l'infrastructure réseau (filaire et WiFi) et prend en compte des objets communicants fixes ou mobiles, des acteurs mobiles équipés de dispositifs, d'acteurs fixes ayant des rôles particuliers utilisant des équipements appropriés. Sur le campus nous avons la possibilité de placer des étiquettes RFID, des bornes RFID fixes et d'utiliser des ordinateurs portables, des TabletPC ou des PDA équipés de cartes WiFi et de lecteurs d'étiquettes RFID. La plateforme IMERA, a bénéficié du soutien de la Région Rhône-Alpes dans le cadre d'un projet Emergence en collaboration avec deux partenaires universitaires (CLIPS-IMAG et MSH-Alpes) et quatre industriels (France-Télécom R&D, TagProduct, Assetium, ainsi que de HP par le projet CAMPUS Mobile).

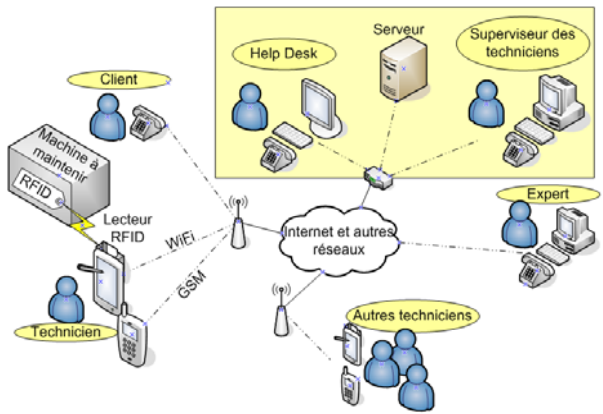

**Figure 1. Plateforme IMERA dans sa déclinaison HMTD pour activités industrielles**

## 3. Etude de cas - Projet HMTD

Le projet HelpMeToDo (HMTD) a pour but d'exploiter des nouveaux moyens de communication mobiles pour le grand public et les professionnels dans toutes les activités nécessitant de l'aide. Les besoins d'information, de formation, d'assistance, d'aide à la maintenance et de dépannage dans des contextes individuels, collectifs, industriels ou grand public sont donc pris en compte. Le projet HMTD vise à étudier cette problématique de façon générique et déclinable dans ces contextes où les contraintes et exigences sont très différentes. L'étude se base sur une exploration des solutions possibles et leur validation lors d'expérimentations « terrain » en contexte réaliste. Dans cet article, nous adressons volontairement les industries à fortes exigences sécuritaires sans s'orienter spécifiquement vers l'aéronautique ou les industries chimiques ou nucléaires.

### 3.1 Contexte
La prise en compte du contexte est incontournable. Pour sa définition nous reprenons celle de Dey [5] qui met en avant 3 aspects (environnement – plateforme – préférences de l'utilisateur) qui sont parfaitement en phase avec nos préoccupations.

### 3.2 Stockage in-situ
Dans les applications mobiles sensibles au contexte comme HMTD il est important d'accéder facilement aux informations décrivant, caractérisant ce contexte. Des objets d'environnement (code-barres, tags divers ou tags RFID) peuvent fournir ces informations statiques et élémentaires permettant ensuite aux dispositifs mobiles leur interprétation et sollicitation à distance (via un réseau) des informations plus complètes et plus dynamiques (pouvant se trouver sur un serveur dans un SGDT :

Système de Gestion de Données techniques). Dans certaines applications mobiles il n'est toutefois pas possible de compter systématiquement sur la disponibilité d'accès réseau. Il faut donc disposer sur place d'informations permettant d'agir et d'accéder aux informations contextualisées. Le stockage in-situ d'informations dynamiques constitue donc une exigence importante pour certaines applications. La technologie RFID constitue une bonne solution à ce problème, car certaines étiquettes RFID peuvent être lues, mais également écrites par les dispositifs mobiles (capacité de stockage de quelques Kbits).

### 3.3 Traçabilité
La traçabilité et l'enregistrement des opérations sont une exigence forte dans des opérations de maintenance et de dépannage d'équipements sensibles. Dans le cadre du projet HMTD il s'agit d'obtenir la traçabilité d'opérations dans le cadre professionnel d'une industrie à risque. On imagine que toutes les pièces de l'équipement à maintenir, les outils (tournevis, pince, marteau et autres outils plus sophistiqués) portent des tags RFID. Les opérations de maintenance se font en mémorisant pour chaque opération la pièce concernée et l'outil utilisé. Par la suite, il est donc possible de retracer les opérations effectuées sur chaque pièce, ainsi que l'utilisation de différents outils. Pour cela les informations collectées seront stockées dans une base de données.

### 3.4 Prescription d'opérations
A l'opposé de la traçabilité se situe le dépannage sous contrôle. Pour assurer la sureté et la robustesse du dépannage il s'agit d'effectuer le travail en suivant un mode prescrit d'opérations. Ce contrôle peut porter d'une part sur l'identité, et donc la qualification du dépanneur, qui doit s'identifier à l'aide de son badge RFID et d'autre part au niveau du processus de dépannage pendant lequel la séquence et les outils utilisés sont contrôlés. Il s'agit donc de prendre connaissance de la séquence d'opérations à effectuer et, pour chacune et dans l'ordre, faire identifier l'outil utilisé comme valide par rapport à ce qui est demandé.

### 3.5 Démarche
La démarche de mise en place d'une application MOCOCO suit la démarche CoCSys (méthodologie de conception de systèmes coopératifs capillaires) proposée par O. Delotte [3]. Suite à la collecte de scénarios partiels tels qu'exprimés par les acteurs impliqués et décrits de manière comparable à [10], on procède à l'élaboration d'un modèle de comportement collaboratif de référence comportant entre autres des arbres de tâches. Ces arbres, suffisamment détaillés, font apparaître les tâches de haut niveau, mais également des tâches cognitives (d'utilisateur), computationnelles et d'interaction faisant intervenir les techniques d'interaction sur lesquelles elles s'appuient. Un référentiel de dispositifs est alors utilisé pour bâtir différentes propositions de configurations de périphériques pouvant satisfaire chacune des tâches. Cette démarche est décrite dans [9]. Une fois les configurations choisies, elles peuvent être modélisées avec IRVO (Interacting with Real and Virtual Objects), qui est un formalisme graphique facilitant la conception, la description et la comparaison des systèmes de Réalité Mixte, mais ne s'occupant pas de la description et de l'implémentation de la partie logicielle [2]. En liaison avec l'arbre de tâches, il permet pour chaque tâche de modéliser l'interaction de l'utilisateur avec les outils et les objets réels et virtuels. Les dispositifs permettant l'échange de données entre les mondes réel et virtuel sont modélisés sous la forme de « transducteurs » : « senseurs » (réel vers virtuel) et « d'effecteurs » (virtuel vers réel).

Dans la suite de ce paragraphe nous montrons les scénarios, deux modèles de tâches, deux configurations de dispositifs dégagées avec la méthode de sélection et la modélisation IRVO des deux activités génériques importantes.

### 3.6 Scénarios

Les scénarios que nous relatons ici portent sur une intervention répondant aux besoins de maintenance sensible et obligatoire nécessitant un respect et un contrôle strict des procédures.

**1.** L'intervenant se connecte à l'application et accède à son environnement ainsi qu'à son planning. Une fois sur le lieu de l'intervention, grâce au tag RFID de la machine concernée, il récupère toutes les caractéristiques de cette dernière : l'historique des interventions, les derniers intervenants, …

**2.** Il lance ensuite le workflow de maintenance correspondant à la machine sur laquelle il veut intervenir.

**3.** La procédure impose à l'intervenant de faire la correspondance entre l'identité de la machine, stockée dans le tag RFID, et celle du workflow. Une fois cette étape validée, la maintenance commence et l'heure de démarrage est sauvegardée dans l'historique de l'intervention.

**4.** Le technicien procède ensuite au démontage, étape suivante du workflow, et utilise ses deux mains, tout en consultant la documentation adaptée, utilisant le média approprié (texte, image, vidéo, son). Voir figure 2.

**Nota :** Lors de chaque étape, le workflow affiche les outils à utiliser, les plans précis des sous-ensembles et pièces à démonter-remonter, et enregistre les actions de l'intervenant. Outils et pièces sont équipés de tags RFID que le technicien doit « taguer » dans l'ordre de la procédure. Cette action conditionne le passage à l'étape suivante.

**5** En cas de rupture de compétence, le technicien collabore à distance avec un expert. Ce dernier accède au contexte, à l'historique de l'intervention et guide l'intervenant via des indications graphiques, orales, textuelles. Voir figure 3.

**6** Si au cours de la maintenance, l'intervenant repère une pièce défectueuse, il peut, via l'étiquette RFID située sur chacune d'elles, lancer un processus de remplacement.

**7** Le remontage est intégré dans la procédure. Chaque pièce et outil sont « tagués » pour vérifier que la maintenance est correctement effectuée.

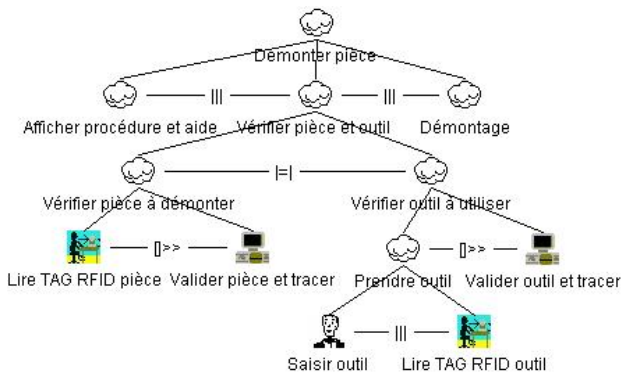

**Figure 2. Arbre de tâche de l'étape 4 exprimé avec CTT**

**8** La procédure se termine par la mémorisation d'un historique de maintenance dans l'étiquette RFID de la machine

### 3.7 Tâches modélisées

A titre d'exemple, les étapes 4 et 5 sont modélisées sur les figures 2 et 3 sous forme d'arbre de tâches CTT faisant apparaître les tâches élémentaires d'interaction en entrées et en sorties ce qui nous permet d'appliquer la démarche d'aide au choix de dispositifs pour l'ordinateur porté (Cf. § 3.8). Pour la figure 2, nous ne détaillons que la sous-tâche « vérifier pièce et outil » illustrant le concept de prescription d'opération (§ 3.4) et de traçabilité (§ 3.3).

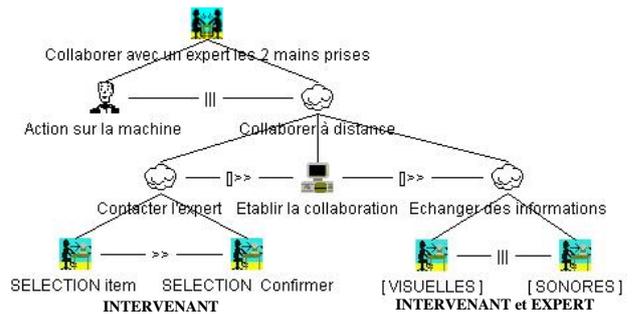

**Figure 3. Arbre de l'étape 5 exprimé avec CTT**

### 3.8 Configurations dégagées

Suite à l'étude des scénarios et à la modélisation des tâches, avec la démarche d'aide au choix de dispositifs pour l'ordinateur porté [9] nous obtenons deux configurations. La première configuration se caractérise par les lunettes à écran opaque intégré, le lecteur RFID dans le creux de la main, le gant numérique pour la sélection d'actions à effectuer et le casque avec micro et écouteur intégrés pour la collaboration et la consultation d'informations sonores. Cette configuration a pour but d'assurer la traçabilité des opérations et de valider la bonne utilisation d'outils lors des séquences d'opérations prescrites. La seconde configuration comporte outre le lecteur RFID, le gant numérique, et le casque avec micro et écouteur intégré, les lunettes « see-through » et sur ces lunettes une caméra pour partager le contexte visuel de l'intervenant avec l'expert distant et pour repérer les tags se trouvant sur la machine. De cette façon le guidage de type réalité mixte est possible.

### 3.9 Modélisation IRVO

L'étape suivante de notre démarche est la modélisation des interactions avec IRVO. Il s'agit dans cette étape de faire une analyse prédictive précise des interactions envisagées avec les différentes configurations dégagées par l'étape précédente d'un point de vue ergonomique afin de les évaluer et de les comparer comme expliqué dans [2]. En particulier, la continuité perceptuelle et cognitive pour les systèmes mixtes mobiles telle que définie dans [4] par les auteurs de la notation ASUR++ peut être analysée sur nos schémas IRVO en appliquant la même méthode. Cette étape permet donc d'affiner les choix des dispositifs d'interaction voire de les remettre en cause. A titre d'exemple nous présentons la modélisation de la tâche « démonter pièce » sans la sous-tâche « vérifier pièce à démonter » de l'étape 4 (Cf. figure 2). La figure 4 présente le modèle IRVO dans la configuration n° 1. L'utilisateur manipule l'outil avec sa main (flèche entre le technicien U et l'outil Tr) qui agit sur la machine à réparer (flèche Tr→Or), ce qui correspond à la sous-tâche « démontage » de la figure 2. En prenant l'outil, le lecteur RFID (modélisé comme un « senseur » S1) placé dans sa paume permet de vérifier que c'est bien le bon outil grâce au tag RFID collé sur le manche. Cette confirmation est présentée à l'utilisateur sous forme d'une indication visuelle (modélisée par un outil virtuel Tv) affichée dans les lunettes à écran opaque (« effecteur » E). La perception visuelle de l'utilisateur est représentée par une flèche Tv→U arrivant sur le canal visuel (V). Cette 2$^{ème}$ boucle action-perception correspond à la sous-tâche « vérifier outil à utiliser » de la figure 2. Enfin, la sous-tâche « Afficher procédure et aide » se réalise par la perception par l'utilisateur de l'objet virtuel « procédure et aide » affiché dans les lunettes opaques.

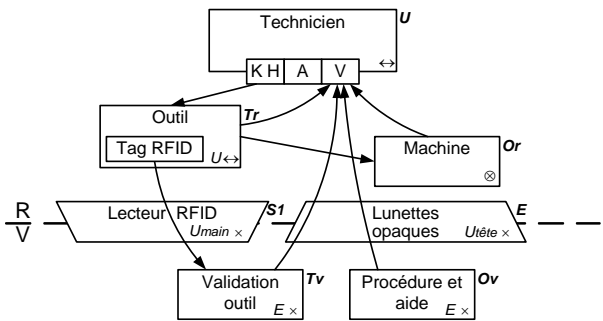

**Figure 4. Modèle IRVO dans le cas de la configuration 1**

La Figure 5 présente le modèle IRVO dans la configuration n° 2. La partie correspondant au lecteur RFID est presque similaire au cas précédent ; cependant, les lunettes « see-through » (effecteur E), permettent d'afficher les informations (Tv) en surimpression sur la scène réelle. Le modèle IRVO fait clairement apparaître dans ce cas la perception augmentée de l'outil réel (Tr) par, d'une part le cadre pointillé (T) entourant l'ensemble des outils réels et virtuels, et d'autre part l'opérateur '+' qui montre la fusion des perceptions visuelles de l'outil réel (Tr) et de l'outil virtuel (Tv). La caméra (senseur S2), fixée aux lunettes, permet de capturer la position de la machine à l'aide de tags visuels et ainsi de pouvoir afficher en surimpression la procédure et l'aide. A nouveau, le modèle IRVO fait apparaître la perception augmentée de la machine réelle grâce au cadre pointillé (O) et à l'opérateur '+'.

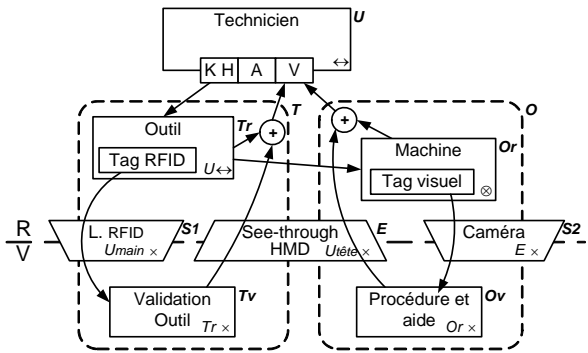

**Figure 5. Modèle IRVO dans le cas de la configuration 2**

Ces deux modélisations IRVO nous permettent, par exemple, de comparer la propriété ergonomique de continuité perceptuelle pour chaque configuration. En effet, dans la figure 4, le technicien doit continuellement regarder 4 sources d'informations (2 réelles et 2 virtuelles) géographiquement dispersés dans l'environnement perceptuel alors que dans la figure 5, via la réalité augmentée, il n'y a plus que deux sources d'informations. La deuxième configuration apparaît donc plus ergonomique réduisant la charge cognitive du technicien.

## 4. Conclusion

Nous avons présenté une étude de cas d'équipement d'un ordinateur porté par des dispositifs d'interaction appropriés aux activités à mener. Nous nous trouvons dans le contexte de la réalité augmentée car nous agissons et communiquons avec les objets et outils tant virtuels que réels. La démarche globale a été présentée ainsi que les résultats obtenus. Des évaluations d'utilisabilité et d'acceptabilité sont en cours avec un support d'évaluation original qui sera bientôt publié. Des applications dans l'industrie sont envisagées, notamment avec notre partenaire Assetium, spécialiste des SGDT (Systèmes de Gestion de données Techniques), très intéressé par leur utilisation pour la contextualisation dans la mobilité.